\documentclass[aps,letterpaper,prd,10pt,onecolumn,showpacs,showkeys,preprintnumbers,superscriptaddress,amsmath,amssymb,nofootinbib,notitlepage,twocolumn]{revtex4-2} 
\usepackage{graphicx, epsfig, amssymb} 
\usepackage{amsmath, amsfonts}
\usepackage{bm} 
\usepackage{color}
\usepackage[usenames]{xcolor}
\usepackage{hyperref}
\usepackage{siunitx}
\usepackage{verbatim}
\usepackage[utf8]{inputenc}
\usepackage{bbm}
\usepackage{booktabs}

\def\ri{{\rm i}}

\def\rd{{\rm d}}
\def\rD{{\rm D}}
\def\tr{{\rm tr}}

\begin{document}
 
\title{Non-Abelian black holes in conformal gravity}

\author{Daniel \surname{Flores-Alfonso}}
\email[]{danflores@unap.cl}
\affiliation{Instituto de Ciencias Exactas y Naturales, Universidad Arturo Prat, Avenida Playa Brava 3256, 1111346, Iquique, Chile}
\affiliation{Facultad de Ciencias, Universidad Arturo Prat, Avenida Arturo Prat Chac\'on 2120, 1110939, Iquique, Chile}


\begin{abstract}
The first black hole solutions of the SU(N) Bach-Yang-Mills equations are presented. Static generalizations breaking spherical symmetry are also constructed. These constitute the first examples in the literature of C-metrics sourced by a Yang-Mills field.
\end{abstract}

\maketitle

\section{Introduction}

In standard gravity, the region outside of astrophysical black holes is uniquely described by the Kerr metric\footnote{Kerr's original work may be consulted in Ref.~\cite{Kerr:1963ud}, while, for discussion of its uniqueness we suggest Ref.~\cite{Heusler:1996}}. To date, 
this prediction of general relativity is consistent with the data found by gravitational wave detectors and large radio telescope arrays\footnote{In 2016, the Laser Interferometer Gravitational-Wave Observatory reported in Ref.~\cite{LIGOScientific:2016aoc} the first black hole binary merger. Soon afterward, in 2019, the Event Horizon Telescope reported in Ref.~\cite{EventHorizonTelescope:2019dse} on the first image of a supermassive black hole.}. However, before their enormous phenomenological success black holes proved to be fascinating mathematical objects and have been studied in an abundance of theories beyond general relativity, including conformal gravity~\cite{Riegert:1984zz,Liu:2012xn}.

Non-vacuum black holes have also found application in astrophysics, e.g., it was recently shown that the plasma in virialized dark matter haloes is able to endow primordial black holes with electric charge~\cite{Araya:2022few}. At the same time, it has also been argued that non-Abelian black holes may be physically relevant in the early stages of the universe. For instance, around phase transitions of unified field theories, see Ref.~\cite{Kleihaus:2016rgf} and references therein. 

In this paper, we find new black holes in conformal gravity coupled to Yang-Mills theory. The non-Abelian fields are chosen to be purely magnetic so that the black holes are able to remain static, cf.~\cite{Heusler:1993cj}. Since the fields are merons the spacetime metric has a different falloff from colored black holes~\cite{Volkov:1989,Bizon:1990sr}. This distinction proves to be crucial as exhibited by the \emph{spin from isospin} effect\footnote{This remarkable phenomenon describes the emergence of fermionic excitations from a purely bosonic gauge theory~\cite{Jackiw:1976xx,Hasenfratz:1976gr,Goldhaber:1976dp}.}, as recently explained in Ref.~\cite{Canfora:2022nso}.

The focus of our work is in conformal gravity, a higher-derivative theory that is renormalizable and so exhibits well-defined ultraviolet behavior~\cite{Adler:1982ri}, which is not the case for general relativity. However, as with most higher-curvature theories, it suffers from the existence of ghosts. Nonetheless, it is quite intriguing that Einstein gravity is, in fact, equivalent to conformal gravity when Neumann boundary conditions are employed~\cite{Miskovic:2009bm,Maldacena:2011mk,Anastasiou:2016jix}. In conformal gravity, the (fourth-order) Bach tensor replaces both the Einstein tensor and the cosmological constant term in the field equations. Straightforward calculations show that every Einstein spacetime is also a solution of conformal gravity. Hence, since the theory enjoys conformal invariance, it follows that every conformally-Einstein metric is Bach-flat. Thus, the significant spacetimes of the theory are those which are inequivalent to Einstein vacua via global conformal transformations.

In the literature, conformal gravity coupled to SU(2) gauge fields has been investigated before, either using numerical approaches or searching for colored black holes as in Refs.~\cite{Brihaye:2009hf,Fan:2014ixa}. However, many more aspects of the theory remain to be explored. Since non-Abelian black holes are much less understood than their (electro)vacuum counterparts we opt for fields of meron type~\cite{deAlfaro:1976qet}. This guarantees that the field is genuinely non-Abelian, as merons cannot possibly exist in Abelian theories. Such a strategy has proven to be fruitful in standard gravity~\cite{Canfora:2012ap}. 

The starting point is to consider an SU(2) hedgehog ansatz for the gauge potential and then to nontrivially embed the system into the SU($N$) theory aligning its internal directions with an irreducible representation of the $\mathfrak{su}$(2) lie algebra occurring within the gauge algebra~\cite{Canfora:2022nso}. The resulting field is physically distinct from the SU(2) system~\cite{Balachandran:1982ty,Balachandran:1983dj,Ayon-Beato:2019tvu}. What is more, the spin from isospin effect is able to discriminate between fields embedded into different gauge groups; see~Ref.~\cite{Canfora:2022nso}.

Since the field ansatz is static it is compatible with a spherically symmetric background. In this way black hole solutions are readily found. We then find that these solutions are amenable to C-metric generalizations. In general relativity, these are the static vacuum solutions with the least amount of symmetry~\cite{Ehlers:1962}. To the best of our knowledge, these are the first self-gravitating Yang-Mills fields on a C-metric background. Through the years, there has been a continued interest in static non-Abelian systems with only axial symmetry, such as black holes or (multi)monopole configurations~\cite{Manton:1977ht,Rebbi:1980yi,Kleihaus:1997ic,Kleihaus:1997ws,Hartmann:2001ic,Radu:2004gu,Herdeiro:2016plq,Gervalle:2022vxs}. Hence, these static axisymmetric solutions could be of interest. 

\section{Spherical Symmetry in Conformal Gravity}

This work focuses primarily on spherically symmetric black holes. In conformal gravity, a Birkhoff theorem was established by Riegert for vacuum and electrovacuum solutions~\cite{Riegert:1984zz}. As such, this result is paramount to us, and we find it convenient to briefly review the Riegert geometry in this preliminary section.

Let us start by writing the Bach-Maxwell equations as
\begin{equation}
    B_{\mu\nu} = 2F_{\mu\alpha}F_{\nu}^{\phantom{\nu}\alpha}-\frac{1}{2}F^2g_{\mu\nu},
\end{equation}
where the Bach tensor is related to the Weyl and Ricci curvature tensors by
\begin{equation}
    B_{\mu\nu} = \left(\nabla^{\beta}\nabla^{\alpha}+\frac{1}{2}R^{\alpha\beta}\right)W_{\alpha\mu\beta\nu}.
\end{equation}
Riegert showed that the general spherically symmetric solution to these equations has background geometry
\begin{subequations} \label{riegert}
\begin{equation} 
  \rd s^2 = -f(r)\rd t^2 +\frac{1}{f(r)}\rd r^2 + r^2\rd\theta^2 + r^2\sin^2\theta\rd\phi^2,
\end{equation}
with
\begin{equation}
    f(r) = a + br + cr^2 + \frac{d}{r}.
\end{equation}
\end{subequations}
For an electric field
\begin{equation} \label{electric}
    F = \frac{q}{r^2}\rd t \wedge \rd r,
\end{equation}
the integration constants must satisfy
\begin{equation} \label{abdq}
     1 - a^2 + 3bd =6q^2.
\end{equation}

The vacuum solution, $q=0$, has been found to possess a form invariance within its conformal class~\cite{Schmidt:1999vp}. By making a coordinate change
\begin{equation}
    r=\frac{\tilde{r}}{1+\alpha \tilde{r}},
\end{equation}
and a conformal transformation $\rd \tilde{s}^2 = \Omega^2\rd s^2$ with
\begin{equation}
    \Omega = 1+\alpha \tilde{r},
\end{equation}
it is found that the resulting metric has the form \eqref{riegert}. The new parameters
\begin{subequations} \label{Schmidt}
    \begin{align}
     \tilde{a} &= a+3\alpha d, \\
     \tilde{b} &= b + 2\alpha a +3\alpha^{2}d, \\
     \tilde{c} &= c + \alpha b + \alpha^{2}a + \alpha^{3}d, \\
     \tilde{d} &= d,
    \end{align}
\end{subequations}
automatically satisfy the corresponding vacuum condition $1 - \tilde{a}^2 + 3\tilde{b}\tilde{d}=0$, as it should since the Bach tensor is a conformal invariant.

Of course, the Schwarzschild metric is a particular case of this geometry. Consequently, the form invariance described above implies that the general vacuum solution is locally equivalent to a Schwarzschild metric via a conformal transformation. This duality, however, is merely local as the conformal factor is singular in some regions of spacetime. To further emphasize the inequivalence of the two spacetimes let us compare the mass of each one; which is a global spacetime property\footnote{A similar situation arises in 2+1 gravity where conserved quantities allow us to distinguish between anti-de Sitter space and the famous Ba\~nados-Teitelboim-Zanelli (BTZ) black hole~\cite{Banados:1992wn,Banados:1992gq,Ayon-Beato:2004ehj,Briceno:2024ddc}.}. The Noether-Wald charge associated with time translation symmetry in conformal gravity is~\cite{Corral:2021xsu}
\begin{equation} \label{mass}
    M = Q\left[\frac{\partial}{\partial t}\right] = \frac{16\pi}{3}\left[ (a-1)b-6cd\right].
\end{equation}
For the Schwarzschild black hole, $f(r)=1-2m/r$, the Noether-Wald mass vanishes in conformal gravity, whereas, the conformally Schwarzschild solution has $M=64\pi\alpha^3m^2$. In fact, any two locally equivalent Riegert spacetimes, cf. Eq.~\eqref{Schmidt}, are globally inequivalent according to the Noether-Wald mass. Lastly, let us remark that the local equivalence between the spherically symmetric vacua of general relativity and conformal gravity does not carry over to the charged case, as the Reissner-Nordstr\"om and Riegert metrics do not share the same conformal class.

\section{Meronic black holes}

There are a number of ways in which Yang-Mills fields are nonunique. For instance, under certain circumstances gauge potentials that are not gauge equivalent may produce the same field strength~\cite{Wu:1975vq,Deser:1976wj}. Phenomena such as these imply non-Abelian black holes are generically not subject to uniqueness theorems---even to the point where some exhibit infinite nonuniqueness, as described in Ref.~\cite{Kleihaus:2016rgf}. However, when systems produce a spherically symmetric energy-momentum tensor then Birkhoff theorems must determine the background geometry. The self-gravitating meron of Ref.~\cite{Canfora:2012ap} is an example of this. In that solution the background has a Reissner-Norsdtr\"om metric with the role of the electric charge played by the Yang-Mills coupling constant. The solutions we present in this section behave similarly.

Let us start by coupling the SU($N$) Yang-Mills theory to conformal gravity according to the following action principle
\begin{equation} \label{action}
 I[g,A]=\int\rd^4x\sqrt{-g}\left(W^2-\frac{1}{2e^2}\tr F^2\right),
\end{equation}
where $F$ the Yang-Mills field strength is defined by $F = \rd A + \frac{1}{2}[A,A]$. The Yang-Mills equations result from variation with respect to the gauge potential $A$
\begin{equation} \label{YM}
    \rD\star F = 0,
\end{equation}
where the gauge covariant derivative is given by
\begin{equation}
    \rD\omega = \rd\omega + [A,\omega].
\end{equation}
Similarly, variation of the action with respect to the metric yields
\begin{equation} \label{Bach}
    B_{\mu\nu} =T_{\mu\nu}= -\frac{1}{2e^2}\tr\left(F_{\mu\alpha}F_{\nu}^{\phantom{\nu}\alpha}-\frac{1}{4}F^2g_{\mu\nu}\right).
\end{equation}

Since we are using matrix-valued differential forms to describe Yang-Mills theory, the gauge potential is written as $A=\ri A^kT_k$ where the $A^k$ are real-valued one-forms and the $T^k$ are matrices representing the gauge algebra. Throughout, we use $N\times N$ traceless Hermitian trace-orthogonal matrices to represent the $\mathfrak{su}$($N$) gauge algebra.

In order to construct our solutions we follow Ref.~\cite{Canfora:2022nso}, where non-Abelian black holes were built from SU($N$) meron sources. Their approach uses a particular description of the SU($N$) group which generalizes the classical Euler-angle parameterization of the SU(2) group. The parameters themselves are referred to as generalized Euler angles. These merons were found to generate spin $s=(N-1)/2$ test fields via the spin from isospin effect. 

The strategy is to first parameterize a meron gauge potential with an SU($N$)-valued scalar field $U$ which maps spacetime points into the generalized Euler angles of SU($N$). The ansatz aligns the meron precisely along the spin-$s$ matrices corresponding to an irreducible representation of $\mathfrak{su}$(2). This interrelation between SU(2) and SU($N$) is what ultimately leads to the generation of higher-spin fields~\cite{Canfora:2022nso}.

Many versions of this strategy have been successful in finding analytical solutions to gravity coupled with nonlinear matter models~\cite{Canfora:2013osa,Ayon-Beato:2015eca,Flores-Alfonso:2020ayc,Ayon-Beato:2019tvu,Alvarez:2020zui,Cacciatori:2021neu,Cacciatori:2022kag,Henriquez-Baez:2022ubu,Corral:2024xfv}. In search of analytical solutions for the Bach-Yang-Mills equations we have opted to use the simplest among them. Our field ansatz is~\cite{deAlfaro:1976qet}
\begin{equation} \label{meronform}
    A = \frac{1}{2} U^{-1}\rd U,
\end{equation}
which despite its appearance has nontrivial field strength. Only in Abelian theories will such potentials be pure gauge, which is to say, their nature is intrinsically non-Abelian.

In order to appropriately align the gauge potential along the irreducible representation of $\mathfrak{su}$(2) we define~\cite{Canfora:2022nso}
\begin{equation} \label{U}
    U = e^{\ri\Phi S_3}e^{\ri\Theta S_2}e^{\ri\Psi S_3},
\end{equation}
where $\Phi$, $\Theta$, and $\Psi$ are Euler angles and the $S_a$ are the standard $N\times N$ spin matrices defined by
\begin{subequations}
\begin{align}
S_{1}=  &  \frac{1}{2}\sum_{j=2}^{N}\sqrt{(j-1)(N-j+1)}(E_{j-1,j}
+E_{j,j-1}),\\
S_{2}=  &  \frac{1}{2\ri}\sum_{j=2}^{N}\sqrt{(j-1)(N-j+1)}(E_{j-1,j}
-E_{j,j-1}),\\
S_{3}=  &  \sum_{j=1}^{N}\biggl(\frac{N+1}{2}-j\biggl)E_{j,j},
\end{align}
\end{subequations}
where $E_{j,k}$ is the matrix with 1 as its $jk$-th entry and 0 elsewhere. These spin matrices satisfy  
\begin{equation}
 [S_a,S_b]=\ri\varepsilon_{abc}S_c,
\end{equation}
and are normalized as follows
\begin{equation} \label{norm}
 \tr(S_aS_b)=\frac{N(N^2-1)}{12}\delta_{ab}.
\end{equation}

A consequence of Eq.~\eqref{U} is that it allows us to write the potential as simply $A=\ri A^aS_a$. Using Schwarzschild coordinates and the map defined by $\Phi=-\phi$, $\Theta = 2\theta$, and $\Psi=\phi$ yields 
\begin{subequations} \label{meron}
    \begin{align}
        A^1 & = -\sin\phi\rd\theta - \cos\theta\sin\theta\cos\phi\rd\phi, \\
        A^2 & = \cos\phi\rd\theta - \cos\theta\sin\theta\sin\phi\rd\phi, \\
        A^3 & = \sin^2\theta\rd\phi,
    \end{align}
\end{subequations}
which indeed satisfy the SU($N$) Yang-Mills equations, Eq. \eqref{YM}. Furthermore, the energy-momentum tensor simplifies to
\begin{equation} \label{meronTmunu}
    T_{\mu\nu} = \frac{N(N^2-1)}{24e^2}\left(F^a_{\phantom{a}\mu\alpha}F_{\phantom{a}\nu}^{a\phantom{\nu}\alpha}-\frac{1}{4}F^a_{\phantom{a}\alpha\beta}F^{a\alpha\beta}g_{\mu\nu}\right).
\end{equation}
where we have used Eq. \eqref{norm}. 

Once Eq. \eqref{meron} is taken into account then Eq. \eqref{meronTmunu} is readily recognized as the energy-momentum tensor of a magnetic monopole, which is not surprising since these merons are known to be locally equivalent, but globally distinct, to Dirac monopoles~\cite{Canfora:2012ap,Canfora:2022nso}. Surely the reader notices a recurring theme within the manuscript.

Direct calculations show that the Bach-Yang-Mills equations are solved by the Riegert metric, \eqref{riegert}, with the integration constants being subject to 
 \begin{equation} \label{cond}
    1 -a^2 +3bd =\frac{N(N{^2-1})}{8e^2}.
 \end{equation}
This is simply Eq. \eqref{abdq} with the role of the electric charge being played by a combination of the color number $N$ and the Yang-Mills coupling constant $e$. In other words, these meronic black holes are more rigid than the electrovacuum solution as they have one less integration constant. Of course, the vacuum solution is recovered in the limit $e^{2}\to\infty$.

Notice that since the non-Abelian black holes characterized by Eq. \eqref{cond} have charged Riegert geometry they are not conformally related to their Einstein-Yang-Mills cousins~\cite{Canfora:2022nso}. The charged Riegert family does not contain the Reissner-Nordstr\"om black hole as a special case, nor are they conformally related modulo reparameterizations of the radial coordinate. Hence, local equivalence between Riegert and Einstein metrics only exists for the neutral metric. 

An alternative way to establish the local inequivalence between our solutions and the ones of Ref.~\cite{Canfora:2022nso} is to compare the relations between conformal gravity and Einstein gravity under Neumann  boundary conditions~\cite{Miskovic:2009bm,Maldacena:2011mk,Anastasiou:2016jix}. Particularly useful to this argument is Sec. 4 of Ref. \cite{Maldacena:2011mk}, where the general Riegert solution is subjected to Neumann boundary conditions. By employing a convenient choice of conformal gauge and coordinate system the Schwarzschild-AdS metric is recovered from imposing the conditions. However, this approach also fails to produce a Reissner-Nordstr\"om metric from the charged Riegert solution for the same reasons as mentioned above.

As mentioned above, the Noether-Wald mass of the vacuum Riegert black hole was calculated in Ref.~\cite{Corral:2021xsu}. Wald's formalism is based on diffeomorphism invariance and builds on the renowned Noether theorem~\cite{Wald:1993nt,Iyer:1994ys}. The method assigns a charge to an on shell conserved current $J$ associated with the vector field $\xi$ generating the symmetry. In general, the Noether-Wald mass depends on the matter content so we should not expect Eq. \eqref{mass} to hold for our solution \emph{a priori}. Nonetheless, in the sequel we show that, in fact, it does.

The starting point is to compute the Noether current as
\begin{equation}
    J^{\mu}=-2\nabla_{\nu}\left(P^{\mu\nu\alpha\beta}\nabla_{\alpha}\xi_{\beta}+2\xi_{\alpha}\nabla_{\beta}P^{\mu\nu\alpha\beta}\right),
\end{equation}
where $P$ is the functional derivative of the Lagrangian with respect to the Riemann tensor. The Poincar\'e lemma then leads to the prepotential $q$ such that $J^{\mu}=\nabla_{\nu}q^{\mu\nu}$. In particular, for conformal gravity we have that
\begin{equation}
   q^{\mu\nu} = -4\left( W^{\mu\nu\alpha\beta}\nabla_{\alpha}\xi_{\beta}-2\xi_{\beta}\nabla_{\alpha}W^{\alpha\beta\mu\nu}  \right).
\end{equation}

When the generator of a diffeomorphism $\xi$ is a Killing vector then the conserved charge is obtained from integrating the prepotential
\begin{equation}
    Q[\xi] = \int_{\Sigma} \frac{1}{2}\epsilon_{\mu\nu\alpha\beta}q^{\mu\nu}\rd x^{\alpha}\wedge \rd x^{\beta}.
\end{equation}
For the Riegert vacuum solution, the Noether charge associated with time translation symmetry is given by Eq. \eqref{mass}, for which integration is carried out on a sphere in the asymptotic boundary by taking the limit $r\to\infty$.

For general non-Abelian black holes the Noether-Wald approach takes into account the matter sector as well for the mass calculation. Notwithstanding, it is very well known that in static systems purely magnetic gauge fields do not contribute to the Noether charge associated with time translation invariance~\cite{Gao:2001ut,Gao:2003ys}. Specifically, notice that for magnetic Yang-Mills fields such as Eq. \eqref{meron} the contribution to the charge vanishes
\begin{equation}
    q^{\mu\nu}_{\text{YM}}\sim F^{a\mu\nu}A^a_{\phantom{a}\alpha}\xi^{\alpha}=0,
\end{equation}
when $\xi=\frac{\partial}{\partial t}$. In other words, for our black hole solutions the mass is indeed also given by Eq. \eqref{mass}. However, since the system is constrained by Eq. \eqref{cond} the result is
 \begin{equation}
    M = -32\pi cd +\frac{16\pi(1-a)}{9d}\left[1-a^2-\frac{N(N^2-1)}{8e^2}\right].
\end{equation}

Lastly, to close this section let us discuss some properties of our solutions. To begin, let us consider the trivial embedding of the Pauli matrices into the SU(N) theory provided by the generators
\begin{subequations}
\begin{align}
\Lambda_{1}=  &  \frac{1}{2}(E_{1,2}+E_{2,1}),\\
\Lambda_{2}=  &  \frac{1}{2\ri}(E_{1,2}-E_{2,1}),\\
\Lambda_{3}=  &  \frac{1}{2}(E_{1,1}-E_{2,2}),
\end{align}
\end{subequations}
which satisfy
\begin{equation}
 [\Lambda_a,\Lambda_b]=\ri\varepsilon_{abc}\Lambda_c.
\end{equation}
Our solutions may be distinguished from similar merons along those directions, $A=\ri A^{a}\Lambda_a$, by noticing that
\begin{equation}
    \frac{\tr\left(S_1^2+S_2^2+S_3^2\right)}{\tr\left(\Lambda_1^2+\Lambda_2^2+\Lambda_3^2\right)} = \frac{1}{6}(N-1)N(N+1),
\end{equation}
which are exactly the famous tetrahedral numbers.
For the case $N=3$, this proportion is 4 and is a very well-known invariant used to distinguish between distinctly embedded configurations~\cite{Balachandran:1982ty,Balachandran:1983dj}. It has also been used together with the baryonic charge to characterize SU(3) $4$-baryons, as in~\cite{Ayon-Beato:2019tvu}.

Next, let us point out that although the energy-momentum tensor of our solutions have the same form as a Dirac monopole they are not equivalent to SU($N$) Wu-Yang monopoles~\cite{Wu:1975es}. To discuss this point further recall that, in conformal gravity, the dual of Eq. \eqref{electric}
\begin{equation}
    F_{\text{D}} = p\sin\theta \rd \theta\wedge \rd \phi,
\end{equation}
leads to a magnetic monopole gravitating on a Riegert metric. The underlying U(1) gauge structure implies that the magnetic charge satisfies the quantization condition $2p=n$, for some integer $n$. For each value of $n$ there exists a different topological space representing the gauge type and infinitely many of them exist. For the SU(2) Wu-Yang monopole this is not the case, as the gauge symmetry implies that $p=n$ instead, and that there is only one gauge type, provided by an underlying product bundle~\cite{Wu:1975es}. As a consequence, it is impossible for our SU(2) meron to be equivalent to an SU(2) Wu-Yang monopole. Upon observation, our solution appears to be locally equivalent to a monopole with $p=1/2$ as this value is permitted by U(1) gauge symmetry. However, this value is decisively not allowed in the SU(2) theory in order for the gauge potential to be globally defined.

This argument carries over to all even values of $N$, as the gauge symmetry characterizing the subbundle underlying our solutions in those cases is also SU(2). However, for odd values of $N$ the gauge subgroup generated by our merons is instead SO(3). Wu-Yang monopoles with SO(3) gauge symmetry do allow for the value $p=1/2$, and two gauge types~\cite{Wu:1975es}. However, since our embedded solutions are characterized by an SO(3) subbundle the type must match that of the SU($N$) theory, which is given by the product bundle. Thus, although SO(3) Wu-Yang monopoles with $p=1/2$ indeed exist they possess a gauge type that is different from that of our merons.

\subsection{C-metric Generalization}

In the first part of this work, spherically symmetric meronic black holes were built in conformal gravity. In what follows, we find static generalizations of these solutions with a background C-metric. Our results complement earlier C-metrics found in conformal gravity sourced by conformal matter~\cite{Meng:2016gyt,Lim:2016lxk}.

In general relativity, the fundamental properties of C-metrics are well understood and they conform close analogs of the uniformly accelerating particle~\cite{Kinnersley:1970zw,Bonnor:1983}. The metric describes two causally separated black holes accelerating away from each other. Generically, they present conical singularities which are interpreted as the cause of the acceleration. Currently, C-metrics are more transparently understood using coordinates developed by Hong and Teo in Ref.~\cite{Hong:2003gx}. For a physical interpretation of the metric employing those coordinates we suggest Ref.~\cite{Griffiths:2006tk}.

Before we continue it is convenient to first rewrite the Riegert metric as
\begin{subequations} \label{riegertxy}
\begin{equation}
   \rd s^2 = \frac{1}{y^2}\left[-F(y)\rd t^2 +\frac{\rd y^2}{F(y)} +\frac{\rd x^2}{G(x)} +G(x)\rd \phi^2\right],
\end{equation}
with
\begin{align}
F(y) &= c+by+ay^2 +dy^3, \\
G(x) &= 1-x^2.
\end{align}
\end{subequations}
The coordinate transformation $y=1/r$ and $x =\cos\theta$ reverts our change recovering Eq. \eqref{riegert}. 

In light of this expression we propose the following C-metric ansatz
\begin{subequations}
\begin{align}
   \rd s^2 &= \frac{1}{\Omega^2}\left[-F(y)\rd t^2 +\frac{\rd y^2}{F(y)} +\frac{\rd x^2}{G(x)} +G(x)\rd \phi^2\right],\\
   \Omega &= y+{\cal A}x,
\end{align}
together with a the gauge potential of the form
\begin{equation}
    A = \lambda U^{-1}\rd U,
\end{equation}
where
\begin{equation} 
    U = e^{-\ri\phi S_3}e^{2\ri\arccos(x) S_2}e^{\ri\phi S_3}.
\end{equation}
The spherically symmetric solution should be recovered once the acceleration parameter ${\cal A}$ vanishes. Notice that along its internal directions, the gauge potential is explicitly given by
    \begin{align}
        A^1 & = \frac{2\lambda\sin\phi}{\sqrt{1-x^2}}\rd x - 2\lambda x \sqrt{1-x^2}\cos\phi\rd\phi, \\
        A^2 & = -\frac{2\lambda\cos\phi}{\sqrt{1-x^2}}\rd x - 2\lambda x \sqrt{1-x^2}\sin\phi\rd\phi, \\
        A^3 & = 2\lambda(1-x^2)\rd\phi. 
    \end{align}
\end{subequations}

It is straightforward to show that the energy-momentum tensor is conserved off shell for the meron. As a consequence, the matter equations decouple from the gravity equations. The former are satisfied only if
\begin{subequations} 
\begin{equation}
    \lambda = \frac{1}{2},
\end{equation}
while, the latter yield
\begin{align}
F(y) &= c +by +ay^2 +dy^3, \\
G(x) &= s +px +lx^2 +qx^3,
\end{align}
with the constraint
\begin{equation}
    l^2-a^2+3(bd-pq) =\frac{2N(N{^2-1})\lambda^2(1-\lambda)^2}{e^2}.
\end{equation}
Notice that the additional integration constants $l$, $p$, $q$, and $s$ can be scaled to match \eqref{riegertxy} whenever ${\cal A}=0$, in order to recover the black hole solution. Furthermore, in the vacuum limit, this C-metric spacetime solves the Bach equations. It will not, however, solve the Einstein equations unless the integration constants are severely constrained.
\end{subequations}

\section{Final Remarks}

This paper's results can be summarized as follows. By considering SU($N$) gauge fields coupled to conformal gravity we have found novel analytical solutions to the Bach-Yang-Mills equations. In the general case, the gauge field is static and gravitates on a C-metric background. Specializing to spherical symmetry produces non-Abelian black holes whose geometry is known from the Birkhoff theorem~\cite{Riegert:1984zz}. For the solutions presented here, the spacetime metric is slightly rigid, i.e., it has one less integration constant than the electrically charged black hole. In our case, the place of the electric charge has been taken by fixed parameters coming from the Yang-Mills theory.

Given that conformal gravity possesses a high degree of freedom, represented by its Weyl symmetry, we have taken advantage of this work to reanalyze the Riegert vacuum solution, whose metric was already known to be conformally related to that of Schwarzschild. Here, we have stressed that this equivalence is merely local and does not carry over to their global structures. To emphasize the inequivalence of the two spacetime we compare their Noether-Wald masses and find that they differ. In any case, this local relationship with Einstein vacua only occurs for neutral Riegert spacetimes, as charged Riegert geometries are not locally related to Reissner-Nordstr\"om black holes by conformal transformations.

The relevance of the solutions provided here is that, for $N>2$, no non-Abelian black holes have been reported in the literature for conformal gravity. In addition, for $N=2$, no meronic black holes were previously known. Furthermore, it seems that no gravitating non-Abelian fields have ever been found with a C-metric spacetime background. Thus, it would be rather interesting to analyze them in the context of standard gravity. Moreover, our results complement previous investigations carried out in Yang-Mills theory coupled to gravity such as those that consider conformal gravity~\cite{Brihaye:2009hf,Fan:2014ixa} and those where the gauge group is SU($N$)~\cite{Shepherd:2015dse,Shepherd:2016ily}.

The approach we have taken here has been to embed $\mathfrak{su}$(2) fields into SU($N$) Yang-Mills theory. This is an arguably natural way to proceed and was commented upon in the renowned work of Belavin, Polyakov, Schwartz, and Tyupkin (BPST)~\cite{Belavin:1975fg}. Notwithstanding, it is lesser known that there exist inequivalent ways to embed the same SU(2) field and that different embeddings lead to physically distinct outcomes~\cite{Wilczek:1976uy,Bitar:1976ji,Ward:1982ft}. For instance, the spin from isospin effect is able to distinguish between meronic solutions embedded along reducible and irreducible representations of $\mathfrak{su}$(2) in the gauge algebra.

To close we comment on possible extensions of our results. Firstly, as mentioned above conformal gravity is closely related to general relativity. However, when nontrivial torsion is considered the relationship between Weyl and Einstein gravity becomes even closer~\cite{Wheeler:2013ora,Wheeler:2018qxx}. Thus, searching for solutions with nontrivial torsion is a possible direction in which our results could, in principle, be extended. Secondly, due to the presence of conformal invariance in our matter content, the solutions presented here may also be subject to analysis in recent emerging research topics, such as those in Refs.~\cite{Alvarez:2022wcj,Anastasiou:2022wjq,Ayon-Beato:2023bzp,Ayon-Beato:2023lrn,Anastasiou:2023sco,Anastasiou:2023oro,Aviles:2024muk,Barrientos:2024umq}. Lastly, our findings could constitute an incentive for further exploration into conformal gravity, complementing established results in standard gravity~\cite{Canfora:2017yio,Giacomini:2019qov,Canfora:2021nca,Canfora:2023bug}.

\section*{Acknowledgments}

For their kind hospitality, the author would like to thank the members and staff of the Departamento de F\'isica of Universidad de Concepci\'on where part of this work was carried out. He also thanks Georgios Anastasiou, Eloy Ay\'on-Beato, Fabrizio Canfora, Crist\'obal Corral, Borja Diez, Gast\'on Giribet, Mokhtar Hassa\"ine, Marcela Lagos, Julio Oliva and Aldo Vera for enlightening discussions. His research is supported by Agencia Nacional de Investigaci\'on y Desarrollo (ANID) under Grant FONDECYT No. 3220083.

\end{document}